\documentclass[pra,aps,twocolumn]{revtex4}
\usepackage{times} 
\usepackage{amsfonts}
\usepackage{amssymb}
\usepackage[dvips]{graphicx}
\usepackage[T1]{fontenc}
\usepackage[latin2]{inputenc}

\usepackage{amsmath}

\newcommand{\ignore}[1]{}

\def\duzomniejsze{<\kern-.7mm<}
\def\duzowieksze{>\kern-.7mm>}

\def\textbf#1{{\bf #1}}
\def\beq{\begin{equation}}
\def\eeq{\end{equation}}
\def\be{\begin{equation}}
\def\ee{\end{equation}}
\def\ben{\begin{eqnarray}}
\def\een{\end{eqnarray}}
\def\beqa{\begin{eqnarray}}
\def\eeqa{\end{eqnarray}}
\def\eea{\end{array}}
\def\bea{\begin{array}}
\newcommand{\bei}{\begin{itemize}}
\newcommand{\eei}{\end{itemize}}
\newcommand{\bee}{\begin{enumerate}}
\newcommand{\eee}{\end{enumerate}}

\def\>{\rangle}
\def\<{\langle}
\def\blacksquare{\vrule height 4pt width 3pt depth2pt}

\def\ot{\otimes}

\def\dt#1{{{\kern -.0mm\rm d}}#1\,}

\def\sigalpe{{\sigma_\alpha'}^{\kern-.7mm E}}
\def\sigalpb{{\sigma_\alpha'}^{\kern-.7mm B}}

\def\ep{\epsilon}
\def\ec{E_C}
\def\ef{E_F}

\newtheorem{lemma}{Lemma}

\newtheorem{theorem}{Theorem}
\newtheorem{proposition}{Proposition}
\newtheorem{definition}{Definition}
\newtheorem{conjecture}{Conjecture}
\newtheorem{fact}{Fact}

\def\bep{\begin{proposition}}
\def\eep{\end{proposition}}
\def\bel{\begin{lemma}}
\def\eel{\end{lemma}}

\def\bet{\begin{theorem}}
\def\eet{\end{theorem}}
\def\bed{\begin{definition}}
\def\eed{\end{definition}}
\def\bef{\begin{fact}}
\def\eef{\end{fact}}

\def\bec{\begin{conjecture}}
\def\eec{\end{conjecture}}

\begin{document}

\title{Remarks on the equivalence of full additivity and monotonicity for the entanglement cost}

\author{F.G.S.L. Brand\~{a}o$^{\dag}$, M. Horodecki$^{*}$, M. B. Plenio$^{\dag}$ and S. Virmani$^{\dag}$}
\affiliation{$^{*}$ Institute of Theoretical Physics and
Astrophysics, University of Gda{\'n}sk, Poland \\$^{\dag}$ QOLS,
Blackett Laboratory, Imperial College London, Prince Consort Road,
London SW7 2BW, UK, \\ $^{\dag}$ Institute for Mathematical
Sciences, Imperial College London, 53 Exhibition Road, London SW7
2BW}







\date{\today}

\begin{abstract}
We analyse the relationship between the full additivity of the
entanglement cost and its full monotonicity under local operations
and classical communication. We show that the two properties are
equivalent for the entanglement cost. The proof works for the
regularization of any convex, subadditive, and asymptotically
continuous entanglement monotone, and hence also applies to the
asymptotic relative entropy of entanglement.
\end{abstract}
\maketitle

\noindent {\bf \em Introduction --} Entanglement is the key
resource in many quantum information processing protocols. In
view of its central character it is of particular interest
to be able to quantify entanglement. With this aim in mind,
basic properties of so-called entanglement measures have
been identified and studied in the literature in some detail
(see \cite{Plenio V 07,Horodecki 01,Horodecki H01} for some
recent overviews).

The detailed character of entanglement and its quantification as a
resource depends on the constraints that are being imposed on the
set of available operations. In a communication setting where two
spatially separated parties aim to manipulate a joint quantum state
it is natural to restrict attention to local quantum operations and
classical communication (LOCC). In this case separable states are
freely available while non-separable states, which cannot be
prepared by LOCC alone, attain the status of a resource that may
achieve a task more efficiently than is possible by classical means
\cite{Masanes 05}. The preparation of non-separable states by means
of LOCC carries a cost in terms of pure singlet states. This gives
rise to the concept of entanglement cost $E_C$. If we denote a
general trace preserving LOCC operation by $\Psi$, and write
$\Phi(K)$ for the density operator corresponding to the maximally
entangled statevector in $K$ dimensions, i.e.\
$\Phi(K)=|\psi_K^+\rangle\langle\psi_K^+|$ where
$|\psi_K^+\rangle=\sum_{j=1}^{K}|jj\rangle/\sqrt{K}$,
then the entanglement cost may be defined formally as
\begin{equation}
    E_C(\rho) = \inf\left\{ r\!:\!\! \lim_{n\rightarrow\infty}
    \left[\inf_{\Psi} D(\rho^{\otimes n},\Psi(\Phi(2^{rn})))\right] = 0 \right\}
    \label{formal}
\end{equation}
where D$(\sigma,\eta)$ is a suitable measure of distance
\cite{Rains 97,Audenaert PE 03,Plenio V 07}. Clearly, the
computation of this infimum is extraordinarily complicated
and it is therefore fortunate that $E_C$ is closely related
to the somewhat more easily computable entanglement of formation
\cite{Bennett DSW 96,Wootters 98}
\be
E_F(\rho) = \min_{\sum_i p_i|\psi_i\rangle\langle\psi_i|=\rho}
\sum_i p_i E_F(|\psi_i\rangle)
\ee
via the limit \cite{Hayden HT 01} \be E_C(\rho) =
\lim_{n\rightarrow\infty} \frac{E_F(\rho^{\otimes n})}{n}.
\label{regular} \ee This equivalence is true for any distance
measure D$(\sigma,\eta)$ that is equivalent to the trace norm with
sufficiently weak dependance upon dimension \cite{Hayden HT 01}.
Here we have defined $E_F$ for pure states as $E_F(|\psi_i\rangle) =
S(tr_A[|\psi_i\rangle\langle\psi_i|])$, i.e. the entropy of
entanglement \cite{Bennett BPS 96} with the von Neumann entropy
$S(\rho)=-$tr$[\rho\log_2\rho]$.

Owing to its operational definition, the entanglement cost suggests
itself as a good entanglement measure. A fundamental property which
an entanglement measure should satisfy (apart from vanishing on
separable states) is monotonicity under local operations and
classical communication (LOCC): \be E(\rho)\geq E(\Lambda(\rho)) \ee
for any LOCC operation  $\Lambda$ which may include mappings to a
larger Hilbert space. It is not hard to check that $E_C$ satisfies
this condition. However, many measures satisfy the somewhat stronger
condition, called full monotonicity, of the {\em non-increase on
average} under LOCC, ie
\begin{equation}
    E(\rho) \ge \sum_i p_i E(\rho_i)
    \label{monotonicity}
\end{equation}
where, in a LOCC protocol applied to state $\rho$, the state
$\rho_i$ with label $i$ is obtained with probability $p_i$. While
entanglement of formation  satisfies this stronger monotonicity, it
is not known whether the entanglement cost does.

Another nontrivial property, that may be shared by some entanglement
measures is full additivity.
A potential entanglement quantifier is additive if it satisfies
$E(\rho\otimes\rho)=2E(\rho)$ for all $\rho$ and it is fully
additive if it satisfies
\begin{equation}
        E(\sigma\otimes\rho) = E(\sigma)+E(\rho)
\end{equation}
for all $\sigma$ and $\rho$. Neither additivity nor full additivity
are trivially satisfied for entanglement quantifiers. The entropy of
entanglement, the so-called squashed entanglement \cite{Christandl W
04}, and the relative entropy of entanglement with reversed
arguments \cite{Eisert AP 03} are fully additive. However the
relative entropy of entanglement is not additive \cite{Vollbrecht W
01}, and if NPPT-bound entangled states exist \cite{DiVincenzo SSTT
00,dur CLB 00} then the distillable entanglement would be
super-additive \footnote{This was first noted in Ref. \cite{Shor ST
01}, where it was shown that if a particular NPPT Werner state is
undistillable, then the distillable entanglement would be
super-additive. In Refs. \cite{Eggeling VWW 01, Vollbrecht W 02}, in
turn, the implication was fully established, as it was shown that
any NPPT state becomes distillable when assisted by a PPT entangled
state.}. It is an unsolved open question whether the entanglement of
formation is additive or not \cite{Shor 04, Matsumoto SW 04,
Audenaert B 04}. Note that the regularisation of an entanglement
measure, i.e. a limit as in eq. (\ref{regular}) is automatically
additive but not necessarily fully additive. Thus the entanglement
cost and the regularized relative entropy of entanglement
\cite{Audenaert EJPVD 01} are additive but it is unknown whether
they are also fully additive.

In this paper we will demonstrate that the two properties of full
monotonicity and full additivity are in fact equivalent for the
entanglement cost. The proof is quite general, and applies to the
regularization of any convex, subadditive, and asymptotically
continuous entanglement monotone, and hence also applies to the
asymptotic relative entropy of entanglement.

To begin with, we recall a useful characterization of the full
monotonicity of convex functions that was established in
\cite{MH2004-mono}.

{\bf Property 1} {\em
A convex function $E$ is full monotone if and only if
it satisfies the following conditions:
\bei
\item {\tt LUI} It is invariant under local unitary operations.
\item {\tt FLAGS} It satisfies
\be
\label{flags}
E(\sum_i p_i  \rho_i \ot |i\>\<i|) = \sum_ip_i E(\rho_i)
\ee
\eei
where  $|i\>$ are local orthogonal flags, $\sum_i p_i=1$,
$\rho_i$ are arbitrary states. }

As the entanglement cost is trivially invariant under local unitary
operations, and is known to be convex \cite{Donald HR} it will
therefore be our aim in the following to  look for  the equivalence
of {\em full additivity} with the FLAGS condition eq. (\ref{flags}).

We will establish that the FLAGS condition is true if and only if
$E_C$ is fully additive, thereby implying that full monotonicity is
true if and only if $E_C$ is fully additive:

\bep $\ec$ satisfies \be \label{eq7} \ec(p \rho\ot \mathbb{P}_0 +
(1-p) \sigma\ot \mathbb{P}_1)= p \ec(\rho)+ (1-p)\ec(\sigma) \ee for
any $\rho$ and $\sigma$ if and only if $\ec$ is additive. Here
$\mathbb{P}_i=|i\rangle\langle i|$ are local orthogonal flags. \eep

\noindent
{\bf Proof --} Denoting by ${\cal S}$
the symmetrised tensor product, e.g. ${\cal
S}(\rho\otimes\sigma)=\rho\otimes\sigma + \sigma\otimes\rho$ then we
find \ben &&\hspace*{-0.65cm} \ec(p \rho\ot \mathbb{P}_0 + (1-p)
\sigma\ot \mathbb{P}_1)
\nonumber \\
&&\hspace*{-0.2cm}=\lim_n {1\over n} \ef\left((p \rho\ot
\mathbb{P}_0 +(1-p) \sigma\ot \mathbb{P}_1)^{\ot n}\right)
\nonumber \\
&&\hspace*{-0.2cm}= \lim_n {1\over n} \ef\!\left({(1-\ep){\tilde
\rho}_{typ} + \ep \tilde \rho}_{atyp} \right) \label{eq8} \een where
we define the `typical part' ${\tilde\rho}_{typ}$ as the {\it
normalized} state proportional to $\sum_{k=\lfloor
pn(1-\delta)\rfloor}^{\lceil pn(1+\delta) \rceil} {\cal S}((\rho\ot
\mathbb{P}_0)^{\ot k} \ot (\sigma\ot \mathbb{P}_1)^{\ot (n-k)})$.
That we are free to write the state in this way is implied by strong
typicality for classical probability distributions, which asserts
that for any $\delta >0$ and $\epsilon
>0$ there exists an $n$ such that
sequences with frequencies of $\rho$ in the interval $[pn
(1-\delta),pn (1+\delta)]$ occur with total probability at least
$1-\ep$ \cite{Cover T}. The remaining term $\tilde{\rho}_{atyp}$ is
defined by eqn. (\ref{eq8}). Note that we assert that the weight of
the typical part is exactly $1-\epsilon$, which we are free to do as
any excess weight can be moved into the definition of
$\tilde{\rho}_{atyp}$.

In $\tilde{\rho}_{typ}$ the local flags give us information about
exactly which of the $n$ copies of our basic two-party system are in
state $\rho$, and which are in state $\sigma$. The typicality
condition ensures that $\rho$ will each occur at least $\lfloor
np(1-\delta)\rfloor$ times, and $\sigma$ will occur at least
$\lfloor n(1-p)(1-\delta)\rfloor$ times. Hence we are free to
perform local unitaries to transform $\tilde{\rho}_{typ}$ to a {\it
sorted} normalized state, $\tilde{\rho}^s_{typ}$, that is
proportional to:
\begin{eqnarray}
\tilde{\rho}^s_{typ} = (\rho\ot \mathbb{P}_0)^{\ot \lfloor
np(1-\delta)\rfloor} \ot (\sigma\ot \mathbb{P}_1)^{\ot \lfloor
n(1-p)(1-\delta)\rfloor} \otimes \tilde{\omega}_{rem} \nonumber
\end{eqnarray}
i.e. we ensure that in $\tilde{\rho}^s_{typ}$ the first
$np(1-\delta)$ copies are in state $\rho$, and the next
$n(1-p)(1-\delta)$ copies are in the states $\sigma$, with the
remaining statistical fluctuations shifted to the last $(n\delta +1)
\pm 1$ copies represented by the remainder state
$\tilde{\omega}_{rem}$. Note that in this process one can ensure
that all local information in the flags will be preserved, so that
in particular one retains the ability to determine precisely whether
one is in the typical or atypical part of the entire state.

Now we use the fact that $\ef$ satisfies the {\tt FLAGS} condition
to find \ben &&\hspace*{-0.65cm} \ec(p\rho\ot \mathbb{P}_0 + (1-p)
\sigma\ot \mathbb{P}_1)
\nonumber \\
&&\hspace*{-0.2cm}= \lim_n {1\over n} \left[\ep\ef\!(\rho_{atyp})+
(1-\ep)\ef(\rho^s_{typ})\right] \een where ${\rho}^s_{typ}$
is the state obtained when we remove all flags from
$\tilde{\rho}^s_{typ}$ (in general the removal of a $\tilde{}\,$
will always signify the removal of any flags). Now we may use the
asymptotic continuity \cite{Nielsen-cont,Donald HR} of the
entanglement of formation to conclude that
\begin{eqnarray} {1 \over
n}\left|\ef\left(\left[p\rho\ot \mathbb{P}_0 +(1-p)\sigma\ot
\mathbb{P}_1\right]^{\ot n}\right) - (1-\ep)\ef(\rho^s_{typ})\right|
\nonumber \\ \le 2\ep C\log d\nonumber
\end{eqnarray}
%
where $C$ is a constant, $d$ is the dimension of a single system
(half of a pair). Now we may use the monotonicity and subadditivity
of the entanglement of formation \be \ef(X)\leq \ef(X \ot Y)\leq
\ef(X)+\ef(Y) \label{monadd} \ee to conclude that
\begin{eqnarray} && \ef(\rho^{\ot \lfloor np(1-\delta)\rfloor} \ot \sigma^{\ot
\lfloor n(1-p)(1-\delta)\rfloor})\leq \ef(\rho^s_{typ}) \nonumber
\\ \leq && \ef(\rho^{\ot \lfloor np(1-\delta)\rfloor} \ot \sigma^{\ot
\lfloor n(1-p)(1-\delta)\rfloor}) +\ef(\omega_{rem}) \nonumber
\end{eqnarray}
and hence the triangular inequality gives \ben && {1 \over
n}|\ef\left(\left[p\rho\ot \mathbb{P}_0 +(1-p)\sigma\ot
\mathbb{P}_1\right]^{\ot n}\right) \nonumber \\ && -
(1-\ep)\ef(\rho^{\ot \lfloor np(1-\delta)\rfloor} \ot \sigma^{\ot
\lfloor n(1-p)(1-\delta)\rfloor})|
\nonumber\\
&& \le 2\ep C\log d\nonumber + 4\delta\log d. \een
Now we let $p_i$ be a rational probability that approximates $p$
closely, i.e. the difference $\ep_i := p-p_i$ is a small real
number. Then there exists an integer $m_i$ such that both $m_ip_i$
and $m_i(1-p_i)$ are integers. Inserting these definitions into the
previous inequality gives:
\ben && {1 \over nm_i}|\ef\left(\left[p\rho\ot \mathbb{P}_0
+(1-p)\sigma\ot \mathbb{P}_1\right]^{\ot nm_i}\right) \nonumber \\
&& - (1-\ep)\ef(\rho^{\ot \lfloor nm_ip_i(1-\delta) +
nm_i\ep_i(1-\delta)\rfloor} \ot \nonumber \\ && \sigma^{\ot \lfloor
nm_i(1-p_i)(1-\delta) - nm_i\ep_i(1-\delta)\rfloor})|
\nonumber\\
&& \le 2\ep C\log d\nonumber + 4\delta\log d. \een
The state in the middle lines of this inequality can be written as
(in somewhat loose terms in order not to obscure the argument):
\ben && \rho^{\ot \lfloor nm_ip_i(1-\delta) +
nm_i\ep_i(1-\delta)\rfloor} \ot \sigma^{\ot \lfloor
nm_i(1-p_i)(1-\delta) - nm_i\ep_i(1-\delta)\rfloor} \nonumber \\
&& \simeq (\rho^{\otimes m_ip_i})^{\otimes n(1-\delta)}
\ot (\sigma^{\otimes m_i(1-p_i)})^{\otimes n(1-\delta)} \nonumber \\
&& \ot \rho^{\ot nm_i\ep_i(1-\delta) \pm 1} \ot \sigma^{\ot
 - nm_i\ep_i(1-\delta) \pm 1} \nonumber \een
Hence a second application of inequality (\ref{monadd}) together
with the triangular inequality gives:
\ben && {1 \over nm_i}|\ef\left(\left[p\rho\ot \mathbb{P}_0
+(1-p)\sigma\ot \mathbb{P}_1\right]^{\ot nm_i}\right) \nonumber \\
&& - (1-\ep)\ef((\rho^{\ot m_ip_i} )^{\ot n(1-\delta)} \nonumber \\
&& \ot (\sigma^{\ot m_i(1-p_i)} )^{\ot n(1-\delta)})|
\nonumber\\
&& \le 2\ep C\log d\nonumber + 4\delta\log d + 2\ep_i(1-\delta)+
2\log(d)/nm_i.\een
Now sending $\ep\to 0$, $\delta\to 0$, and $n \rightarrow \infty$ we
obtain
\ben && |\ec\left(\left[p\rho\ot \mathbb{P}_0
+(1-p)\sigma\ot \mathbb{P}_1\right]\right) \nonumber \\
&& - {1 \over m_i}\ec(\rho^{\ot m_ip_i} \ot \sigma^{\ot m_i(1-p_i)}
)|
\nonumber\\
&& \le  2\ep_i.  \label{eq: main} \een
This inequality is the key to obtaining the proposition. In this
equation if we pick $m_i=2,p=p_i=1/2$, then we find that $\ep_i=0$
and recover the equality:
\be \ec({1\over 2} \rho\ot |0\>\<0| +{1\over 2} \sigma\ot |1\>\<1|)=
{1\over 2} \ec(\rho\ot \sigma) \label{eq:half} \ee
By comparing this equality to the FLAGS condition we see that if
FLAGS is true for $\ec$, then $\ec$ is fully additive. On the other
hand, if $\ec$ is fully additive, then equation (\ref{eq: main})
becomes:
\ben |\ec\left(\left[p\rho\ot \mathbb{P}_0 +(1-p)\sigma\ot
\mathbb{P}_1\right]\right) & - p_i\ec(\rho) - (1-p_i) \ec(\sigma)|
\nonumber\\
&\le  2\ep_i, \nonumber \een
in which case sending $p_i \rightarrow p$ and hence $\ep_i
\rightarrow 0$ establishes the FLAGS condition for all $p$. Hence
fully additivity of $\ec$ also implies full monotonicity.
\blacksquare\\

Finally let us note that we have used the following properties of
$E_F$ and $E_C$: 1) full monotonicity of $E_F$, 2) asymptotic
continuity of $E_F$, 3) sub-additivity of $E_F$, 4) convexity of
$E_C$. The asymptotic continuity follows from convexity and
subadditivity of $E_F$. We thus obtain the following proposition:
\bep Suppose that a function $E$ that is convex, subadditive, fully
monotone under LOCC and asymptotically continuous.
Then full monotonicity and full additivity of $E^\infty$ are
equivalent. \eep

In particular this demonstrates that the asymptotic relative entropy
of entanglement is fully additive iff it is fully monotone.

{\bf Summary and Conclusions --} We have considered the relationship
between the full additivity and full monotonicity of the
entanglement cost. We have established that the two properties are
equivalent for the regularization of any convex, subadditive, and
asymptotically continuous entanglement monotone, including the
entanglement cost.

\medskip

\acknowledgements We acknowledge support by the EPSRC QIP-IRC
(www.qipirc.org), the EU Integrated Projects Qubit Applications
(QAP) and Scalable Quantum Computing ith Light and Atoms (SCALA) and
the Royal Commission for the Exhibition of 1851. MBP holds a Royal
Society Wolfson Research Merit Award. MH is supported by Polish
Ministry of Scientific Research and Information Technology under the
(solicited) grant no. PBZ-MIN-008/P03/2003. FGSLB is supported by the
Brazilian agency Conselho Nacional de Desenvolvimento Cient\'ifico e
Tecnol\'ogico (CNPq).


\end{document}